# ACCESS POINTS PLACEMENT
# AS A DISCRETE OPTIMIZATION PROBLEM

*Lev A. Kazakovtsev[1]*,
*Siberian Federal University*
Krasnoyarsk, Russia, levk@bk.ru

**ABSTRACT**

In this work, we consider a method of searching of the direction of a wireless network development (the places of new access points or base stations etc.) optimized with criteria of coverage of important territories and minimum cost of equipment and additional needed infrastructure which does not need the execution of special field testings and determination of the exact geometry of elements of RF-propagation medium and their RF absorbing properties but takes into account the minimum accessible information obtained from built-in measuring instruments of wireless hardware and approximate data of the medium elements shape. The problem of search of a disposition and types of the infrastructure elements of the growing network is formulated as a multicriteria discrete constrained optimization problem solvable with variant probability method [1]. The problem of a medium RF-propagation properties modeling is also formulated and solved as a discrete optimization task.

## 1.INTRODUCTION

The sparsely populated territories with underdeveloped infrastructure need a special approach for solving the problem of supply with the telecommunication facilities. The uninterrupted rapid growth of the telecommunication networks including qualitative increasing of data channels bandwidth and quality of service is mostly typical for the megapolises and densely populated territories is least apparent in remote and sparsely populated regions where the networks are nevertheless apparent. On the one hand, the absence of any infrastructure (not only telecommunication one but also traditional telephone lines, power transmission lines and roads along which the telecommunication cables infrastructure are usually laid) embarrasses the development of the traditional cable infrastructure. The implementation of ADSL technology using the existing cable infrastructure is also impossible. On the other hand, the necessity to supply that territories with the qualitative network access is dictated by the standards (written or de-facto ones) of the office work and business processes of various business, governmental, educational and research organizations. Furthermore, the remote territories need a qualitative communications which makes them 'nearer' to the 'big earth' even more than the central regions because this is the only way to establish contacts with the partners, to realize centralized management and even to make simple things like bank transfer. In conditions of the Northern Siberia, the existence of any serious cable infrastructure is often absolutely impossible even within small territories (for example, territory of a settlement) because of permafrost, bogs, moving soil. By the way, this problem embarrasses also the placing of wireless base stations, nevertheless, only the wireless technologies have any the serious perspectives at these territories.

Certainly, the satellite technologies allow to establish the Internet access, telephone connection and everything else at every point but they are expensive. That is why, the local area networks (often covering large territories) are usually developed with WiFi technology (often, in addition to VSAT equipment as a gateway to WAN).

Often, such networks are developed chaotic as far as an exigency to cover new territories arises. The equipment (especially arctic versions of devices) is not too cheap. The superfluous quantity of access points does not do much good because of interference and frequency sharing problems but this problem is less actual for sparsely populated territories. The execution of the field testings of RF-propagation properties of medium and its geometry need special equipment, skilled personnel and significant expenditure of labor. Meanwhile, the maximum usage of available information (even if it is not precise) of signal and noise levels at different points and RF absorbing

---
[1] Author gratefully acknowledges the ICTP-ITU 2008 School on New Perspectives in Wireless Networking (2008, Trieste, Italy)

properties of medium acquired from the working wireless network hardware is able to help with the search of the optimal configuration of new elements of the growing network. The same is actual for the wireless networks which are developed chaotic at the large built-up territories (for example, reconstructed territories of the former industrial enterprises).

## 2. RELATED WORKS

The considered problem of optimal access points placement is often a subject of the engineers and network administrators in the Internet forums and scientific publications. Some of them offer ready-to use software able to forecast the signal coverage in accordance with the physical properties of the environment elements. Other scientific works offer various methods of search of new access points placement with optimal signal coverage [2, 3] or (come complex case) placement which provides for the best localization of a mobile WLAN equipment position (calculation of its coordinates) within the area of coverage of the network [4, 5]. The prognosis of signal level at each point taking into consideration the path loss of walls, ceilings and other elements is a well known engineering problem [7]

The recent literature proposes and approves many techniques to analyze the indoor and outdoor signal coverage problem and different deterministic and stochastic optimization methods to find the optimal placement.

In [2], author a version of the Nelder-Mead simplex method and a pattern algorithm that considers the minimization of the ratio of covered points in a mesh.

In [3], there is a method of design of large-scale indoor network where placement of the access points guarantees the absence of the gaps. The optimal signal coverage area of an AP is considered as a cylinder and the AP positions are optimized with usage of geometrical schemes.

Using the empirical Motley-Keenan indoor wave propagation model taking into consideration the type of walls and ceilings to calculate the path loss, [6] proposes an AP placement technique for optimal inside signal coverage. Genetic algorithm is used to find a configuration with less value of maximum path loss for each point.

In [8], authors also propose a strongly typed genetic programming to solve access point configuration problem with optimal disposition of the access points providing the best coverage.

In [4], they use genetic algorithms to find the disposition of access points which gives the optimal localization of the mobile device within some area covered by access points. In [5], they propose a ready-to-use system for the localization of such device.

In our case, the optimal coverage is one of the objectives. The second important one is the minimum cost of the wireless equipment and infrastructure needed to make this equipment work. The results of the above methods are hard to interpret when the 'optimal' placement of an access point is situated far from the electric wiring, walls or other constructions where we are able to install the equipment. The transferal of the resulting position of the access point to the nearest possible place may significantly change the forecast of coverage. The erecting work is possible during a very short period of the year in Siberian weather conditions but we do not know when the necessity to change our network configuration arises.

Most of above works are based on the signal propagation model which considers the logarithmic loss of each obstacle (walls, ceilings) from some information tables but often, we do not know even the exact building material of our environment elements and their exact geometry and the real absorption may differ very significant. In [4], they propose to use the least-squares fit on experimental data to determine the real path loss produced by the obstacles. But we may have enough or not enough experimental data which are needed for the least-squares method and it may be difficult to obtain the additional data.

## 3. AVAILABLE INFORMATION

The minimum information we need is the level of the signal and noise in different points. At least, the information of the signal and noise levels at the points where the network is already working is able to acquire from the network equipment (for example, the "iwconfig" utility gives us such information). The minimum information of the environment geometry (the disposition of walls, trees and other objects) is always available from maps, schemes or immediate visual observation. The RF absorbing properties of the environment elements are available from the information tables [8] and able to be defined more exactly if the element is situated between the existing transmitter and receiver or in the Fresnel zone. In complex cases, the absorbing properties can be specified as a solution of the below optimization problem.

Also, we have an information about the possibility and the approximate cost to allocate the new access point ate each concrete place. So, the places where we have an electric wiring (at least, the places with the existing client wireless hardware) are more suitable to set the new access points than outdoor ones distant from the buildings and power supply infrastructure. Moreover, the researcher is able to set some number of possible places of the new access points in accordance to the reasons which are evident for him but hard to be formalized.

### 4. SUPPOSITIONS, SIMPLIFICATIONS

So, we have the information about the approximate geometry of the environment (disposition of the existing equipment, disposition of the obstacles for signal propagation and their presumable RF absorption properties) and the data of signal and noise levels at some points.

To make our calculations simpler, let's divide our area into cells. As a rule, the real problems do not allow us to ignore the vertical positional relationship of the infrastructure and environment elements. That is why, we should consider our environment as a 3D space. For the simplicity of the statement and representation at the paper, we consider 2-dimensional case.

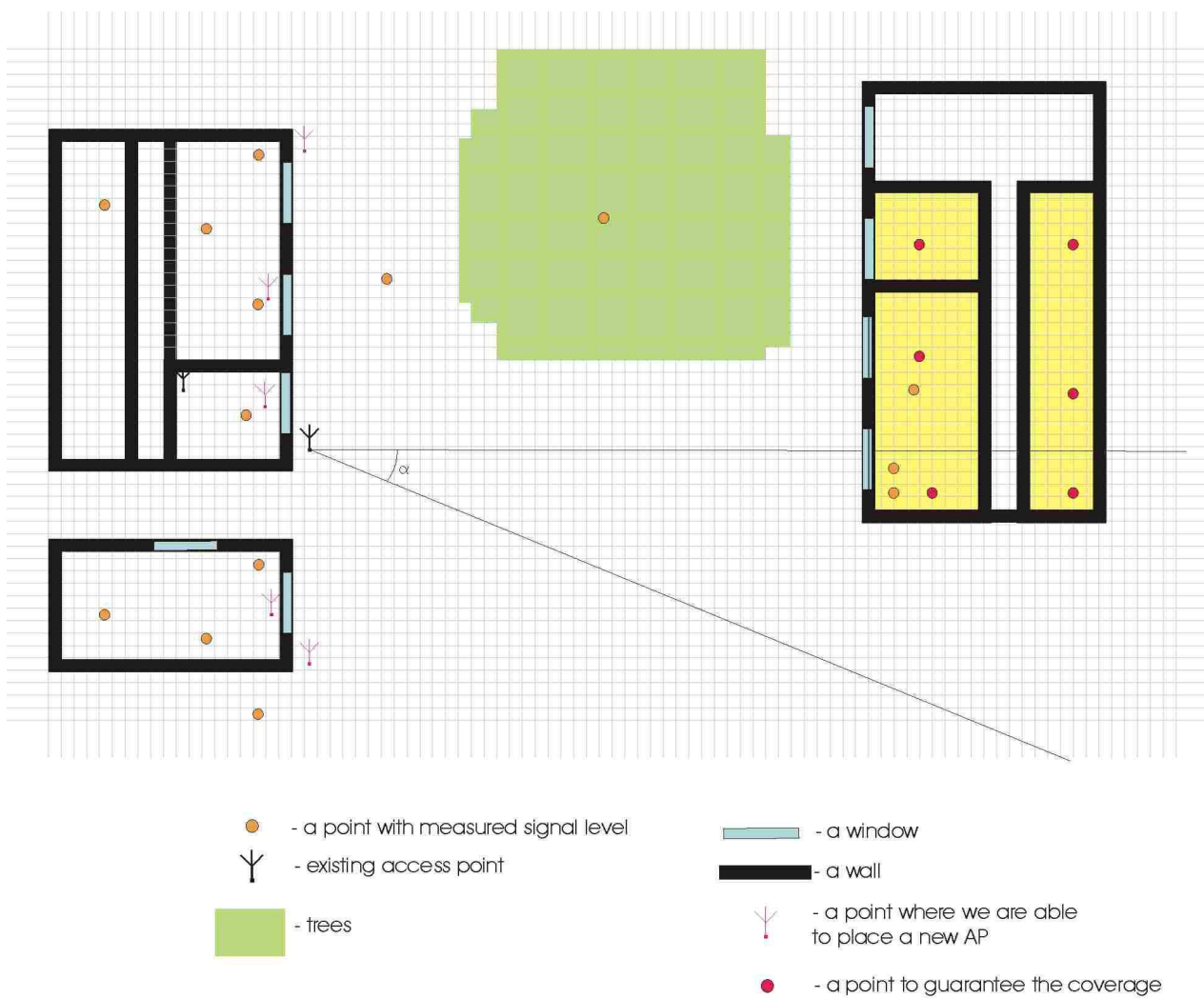

**Figure 1.** Fragment of the scheme of equipment placement and environment elements in discrete coordinates

Let's consider than the equipment is situated in the centers of the cells and the borders of the environment elements (obstacles such as walls, windows, trees etc.) coincide with the borders of any set of the cells. So, we draw the scheme of our environment with the existing and new access points and ,obstacles and existing and perspective zones of coverage of our network as if we mark the elements in the checked writing-book (Fig.1). If we have an information about the signal level at some place then we suppose that this level is equal for the whole cell.

If our network contains directional antennas then we suppose then the signal of such antenna is propagated within the area bounded with an angle (excluding pairs of beam transmission antennas, in this case, we suppose suppose that the area of propagation includes only 2 cells where 2 antennas are situated).

## 5. PROBLEM STATEMENT

Let there be several areas in our scheme which need to become the consistent reception (consistent coverage) areas. Moreover, certain areas are most important, other ones are desirable to supply the consistent coverage but it is not so important. We define the weight coefficient for each of these areas. Besides, we are allowed to demand some minimum bitrate for one or another area. Thus, we have $N_c$ cells where we need to supply the coverage with the weight coefficients v$j$ and minimum bitrates $b_j$, $1 \leq j \leq N_c$.

Let there be $N_p$ points (cells) where we are able to place the access points. For each of these cells, we take into consideration $\underline{N_t}$ types of hardware (differentantennas, amplifiers etc). For each of these places, we know also the approximate cost of infrastructure equipment (such as power supply wires or self-contained power supply, antenna holders etc.) $C_i$, $1 \leq i \leq N_p$ and the cost of each kind of the hardware $C_k$ is also known, $1 \leq k \leq N_c$.

Let's define a matrix $X$ of Boolean variables $x_{ik}$. Setting the value of the variable $x_{ik}$ to 1 means that we decided to place the access point hardware of $k$-th type within the $i$-th possible cell and setting its value $x_{ik}=0$ means that we decided to place there another kind of equipment or not to place any access point within the $i$-th cell.

Our objective is to supply the maximum bitrate at maximum number of cells in accordance with their weight coefficients with minimum expense.

$$b_j(X)v_i \to max,$$
$$\sum_{i=1}^{N_p} \sum_{k=1}^{N_t} (C_i + C_k) x_{ik} \to min,$$
$$\sum_{k=1}^{N_t} x_{ik} \leq 1 \ \forall \ 1 \leq i \leq N_p,$$
$$x_{ik} \in \{0,1\} \ \forall \ 1 \leq i \leq N_p, 1 \leq k \leq N_t,$$
$$b_j \geq b_{minj}.$$

Here, $b_j(X)$ is the maximum possible bitrate at the $j$-th receiving point (though all wireless devices are duplex, for the simplicity, we name he place to be covered by the growing device 'receiving' point in contrast to the access point). The access points are set in accordance with the vector $X$ of Boolean variables, $v_j$ is the weight coefficient of the $j$-th receiving point, $N_p$ is the total number of possible places of access points in the future system, $N_t$ is the number of types of access point equipment (access points or wireless routers equipped with corresponding types of antennas), $b_{minj}$ are minimum guaranteed bitrates needed for some receiving points.

## 6. MATHEMATICAL MODEL

Let's define the RF absorbing properties of each obstacle shown in our scheme. Let $\Pi_l$ be the absorption (in decibels) of the signal passing through the layer of $l$-th obstacle a cell thick. The initia values can be obtained from information tables [8]. For example, for a wall $\Pi_l$= -7 dB, for a window -2 dB, for the forest -... dB for each meter (the value for a cell depends on its size). Below, we consider a method allowing us to define more exactly the value of $\Pi_l$ in accordance with the experimental data.

Then, the signal level (Fig.2) of the *i*-th access point which we can receive at the *j*-th cell can be calculated as

$$P_{rij}(X) = P_{ti} + G_{ti} + G_r - L_{ij}(OBST) - FSL_{ij},$$
$$FSL_{ij} = 40 + 20\log D_{ij},$$
$$P_{rij}(X) = P_{ti} + G_{ti} + G_r - L_{ij}(OBST) - 40 - 20\log D_{ij}.$$

Here, $P_{rij}(X)$ is the signal level (dBm) of the *i*-th access point received at *j*-th receiving point, $G_{ti}$ is the gain of the antenna of the access point (including the loss of all cables), $G_r$ is the gain of the antenna of the receiving point (also, including all cables), $L_{ij}(OBST)$ is the path loss between the *i*-th access points and *j*-th receiving point with the configuration of obstacles (walls, windows, trees etc.) described with a set *OBST*. $D_{ij}$ is the distance (in meters) between the *i*-th AP and the *j*-th receiving point.

A bitrate which is supplied for this signal level is

$$b_{ij}(X) = \begin{cases} 54\,mb/s, P_{rij}(X) - N_j > 66\,dBm, \\ 18\,mb/s, P_{rij}(X) - N_j > 66\,dBm, \\ 1\,mb/s, P_{rij}(X) - N_j > 93\,dBm, \\ 0, P_{rij}(X) - N_j \leq 93 \end{cases}$$

Here, $b_{ij}(X)$ is the maximum available bitrate of the link between the *i*-th access point and the *j*-th receiving point.

Taking into consideration several possible variants of the wireless equipment for each access point, we the signal level is

$$P_{rij} = \sum_{k=1}^{N_t} P_{ti} + \sum_{k=1}^{N_t} G_{ti} x_{ik} + G_r - L_{ij}(OBST) - 40 - 20\log D_{ij}$$

and the available bitrate of the j-th receiving point is
$$(b_j(X)) = \max_i \{b_{ij}(X)\}.$$

To calculate the signal level received at some cell from some access point we should determine which obstacles lay between the access point and the covered cell taking into consideration the Fresnel zone.

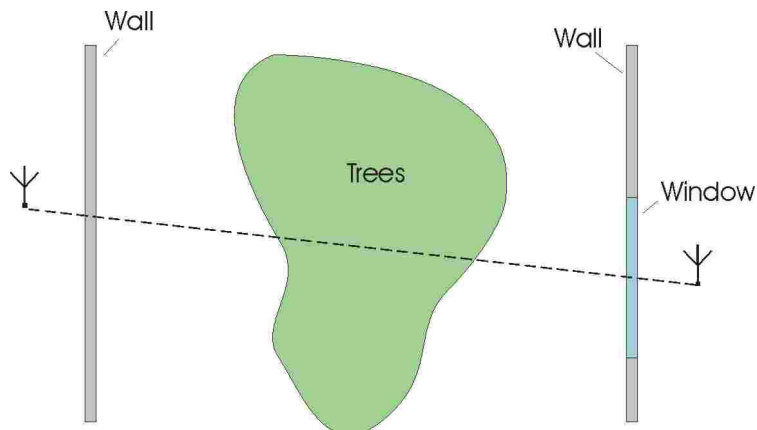

**Figure 2.** Path loss

Let's evaluate the thickness of the Fresnel zone in cells.

To determine the cells which we consider to belong to the direct sight line we implement the line drawing algorithm (which is implemented in the computer graphics software) [11] (Figure 3).

$$R_F = \frac{17.3}{d_{cell}}\sqrt{\frac{1}{(1000 F)}\frac{D_i D_j}{D_i + D_j}} = \frac{0.3531}{d_{cell}}\sqrt{\frac{D_i D_j}{D_i + D_j}}.$$

Here, $R_F$ is the diameter of the Fresnel zone, $d_{cell}$ is the size of a cell in discrete coordinates scheme, $D_i$ is the distance to the $i$-th AP, $D_j$ is the distance to the $j$-th receiving point and F is the frequency (let it be 2.44 GHz).

Let's determine (in cells) the thickness of the Fresnel zone in the middle point. If it does not exceed 1.5 points then we suppose it to be equal to 1 cell and the wave propagation zone lays within the line shown in Figure 3.

$$R_F = \begin{cases} 1, \dfrac{0.3531}{d_{cell}}\sqrt{\dfrac{D_i D_j}{D_i + D_j}} < 1.5, \\ \left|\dfrac{0.3531}{d_{cell}}\sqrt{\dfrac{D_i D_j}{D_i + D_j}}\right|, \dfrac{0.3531}{d_{cell}}\sqrt{\dfrac{D_i D_j}{D_i + D_j}} \geq 1.5 \end{cases}$$

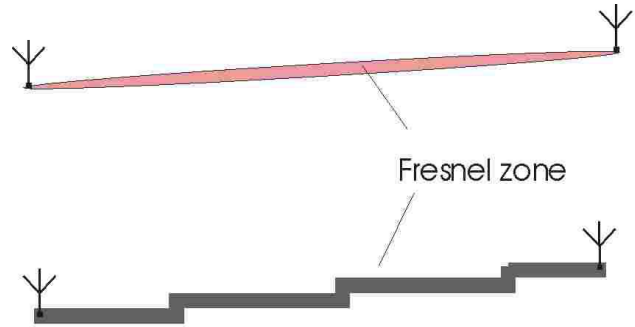

**Figure 3.** Fresnel zone and its presentation in discrete coordinates

Otherwise, for each cell of the line shown in the Figure 3 where the Fresnel zone thickness exceeds 1.5 cells, we 'draw' a line segment normal to the direct sight line (Fig.4) with the length equal to the diameter of the Fresnel zone. Let's investigate what number of cells of this line segment is occupied with the obstacle.

$$r_{Fmij} = \frac{N_{OBSTm}}{N_{Fm}}.$$

Here, $r_{mij}$ is the ratio of the number of cells of the $m$-th line segment occupied with the obstacle ($N_{OBST\,m}$) and the total number of cells in the line segment ($N_{Fm}$).

If the obstacle occupies less than 25% of the line segment cells then we suppose that the absorption of the obstacle at this segment is proportional to the part of the Fresnel zone, if it exceeds 25% then we suppose that the path loss caused by the obstacle is equal to that the obstacle blocks up all the cut of the Fresnel zone.

$$L_{mij} = \begin{cases} L_{OBSTq}, r_{Fmij} \leq 0.25 \wedge OBST_q \cap SEGmij \neq \emptyset, \\ L_{OBSTq}, r_{Fmij} > 0.25 \wedge OBST_q \cap SEGmij \neq \emptyset, \\ 0, OBST_q \cap SEGmij = \emptyset \end{cases}$$

Here, $L_{OBSTq}$ is the path loss caused with the $q$-th obstacle which have the size of 1 cell, $OBST_q$ is the configuration of the $q$-th obstacle (a set of the cells which it occupies), $SEG_{mij}$ is a set of cells of the $m$-th line segment normal to the line from the $i$-th AP to the $j$-th receiving point.

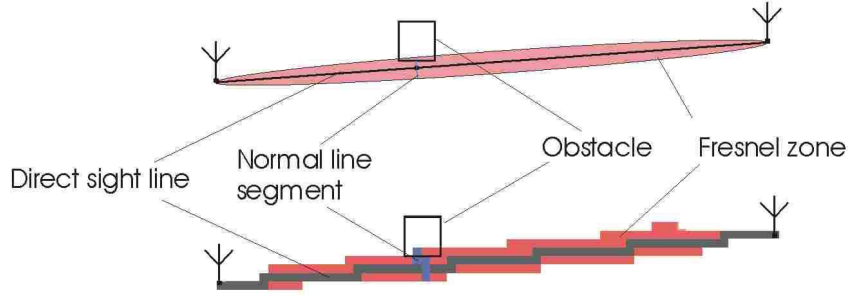

**Figure 4.** Fresnel zone and its presentation in discrete coordinates

If the Fresnel zone is occupied with the obstacle with more than 30% then we suppose that the obstacle absorbs the signal as if it occupies 100% of the Fresnel zone [8].

So, we know the value of loss caused by the obstacles situated between the access point and covered cell.

$$L_{ij} = \sum_{m=1}^{N_{ij}} L_{mij}(OBST)$$

Here, $L_{ij}$ is the total path loss caused by the obstacles situated the $i$-th AP and the $j$-th receiving point.

Thus, the signal level from the $i$-th access point equipped with the hardware of the $k$-th type received within the $j$-th cell is

$$P_{rj}(X) = \sum_{k=1}^{N_t} (P_{ti} + G_{ti}) x_{ik} + G_r - \sum_{m=1}^{N_{ij}} L_{mij}(OBST) - 40 - 20\log N_{ij} d_{cell}$$

The client wireless equipment situated within each of the covered cells is able to establish a connection with any access point but it selects the access point with maximum signal level. The signal level of the 'best' access point received within the $j$-th cell can be calculated as

$$P_{rij}(X^{real}, OBST) = \left\{ \sum_{siz\,8\,k=1}^{N_t} (P_{ti} + G_{ti}) x_{ik} + G_r - \sum_{m=1}^{N_{ij}} L_{mij}(OBST) - 40 - 20\log N_{ij} d_{cell} \right\}.$$

Thus, we have a discrete optimization problem with pseudo-Boolean object functions with constraints having 2 criteria. To solve problems of that kind we have a large experience of implementation of the variant probability algorithm (MIVER) proposed by A.Antamoshkin [1]. The problem of contraction of the 2nd criterion and analysis of of Pareto-optimal set of solutions we use the method proposed in [12]. The algorithm and method have been successfully implemented for a large class of the analogous 2-criteria optimization problems of telecommunication systems [13], so, it is possible to presume its efficiency for our problem. The final step of the algorithm is the comparison of the Pareto-optimal solutions shown at the scheme with the values of possible bitrates of each cell for different values of the cost spent to achieve such coverage.

The maximum bitrate available at the $j$-th point is

$$b_j(X) = \begin{cases} 54\, mb/s, P_{rj}(X) - N_j > 66\, dBm, \\ 18\, mb/s, P_{rj}(X) - N_j > 66\, dBm, \\ 1\, mb/s, P_{rj}(X) - N_j > 93\, dBm, \\ 0, P_{rj}(X) - N_j \leq 93 \end{cases}$$

### 7. MODEL PARAMETERS ADJUSTMENT

We can presume the absorption of each obstacle in our scheme. But usually, we do not know the exact structure of the building materials, exact thickness of the walls and other parameters that causes the significant difference (even in logarithmic scale) between the real values and ones taken from the information tables. Sometimes, there are the obstacles which are not shown in the initial scheme but exert the significant influence upon the signal propagation. Let's solve the problem of accurate definition of the values of absorption for each obstacle and detection of 'invisible' obstacles.

Let the real value of signal level at $j$-th cell be $P_{realj}$ and the calculated value be

$$\left\{ \sum_{k=1}^{N_t} (P_{ti} + G_{ti}) x_{ik} + G_r - \sum_{m=1}^{N_{ij}} L_{mij}(OBST) - 40 - 20 \log N_{ij} d_{cell} \right\}$$

Here, $X^{real}$ is a vector characterizing the configuration of of the current placement of access points ($x_{ik}^{real} = 1$ if there is the $k$-th type of equipment already situated at the $i$-th place).

In this calculation, we take into consideration only the existing access points. Thus, to determine the most adequate values of coefficients ... we should solve the minimization problem

$$\sum_{j=1}^{N_p^{real}} |P_j^{real} - P_{rj}(X^{real}, OBST)| \to min$$

Here, $N_p^{real}$ is the number of the access points which are already functioning.

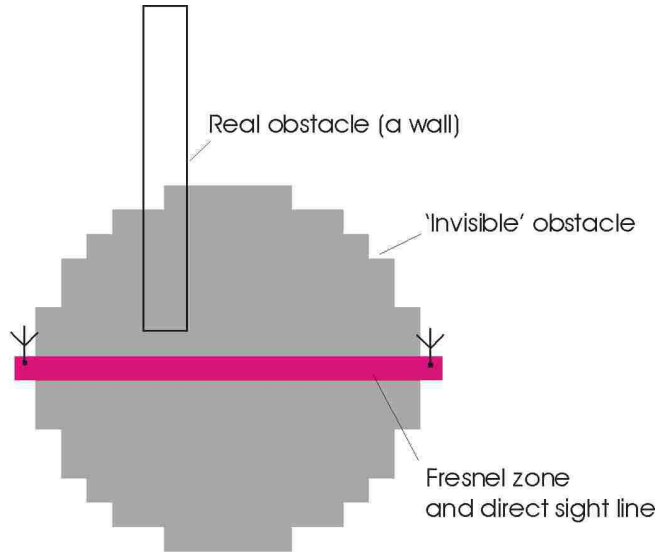

**Figure 5.** Additional 'invisible' obstacle

Here, the optimized variables are ... (values of absorption for eacl                    the absorption values of 'invisible' obstacles. For the simplicity, we assume that these obstacles are situated between the access point and receiving cell and their shape is a sphere or circle with the center between the access point and receiving cell (in the middle, Fig. 5) because we do not know the real shape. We assume that such an invisible obstacle exists when the direct sight line and Fresnel zone is clear but the real measured signal level differs from the calculated one significant (10 dB in our example)

$$L_{ij}^{inv} = \begin{cases} 0, P_{rj}(X^{real}, OBST) - P_j^{real} < 10 \text{dB}, \\ P_{rj}(X^{real}, OBST) - \frac{P_j^{real}}{D_{ij}} l d_{cell}, P_{rj}(X^{real}, OBST) - P_j^{real} \geq 10 dB \end{cases}$$

Here, $L_{ij}^{inv}$ is the path loss caused by each cell of this 'invisible' obstacle situated between the *i*-th AP and *j*-th receiving point.

## 8. RESULTS

We considered the problem of choice of place for additional access points of the growing network and their types as an optimization problem which is based on the adapted model of wave propagation environment.

Though our model takes into consideration only the most important of the propagated signal and environment (practically, only the absorption and noise without detection of its nature), this model allows us to forecast the signal level and bitrate of each perspective area at least as a rough estimation. This method does not need any field testings but it taker into consideration the results of them if they have been performed. Basically, our method coincides with the usual practice of the engineering calculations but it allow to collect all the information about our environment into a single model and find one or more Pareto-optimal solutions taking into account all the information of the environment and cost of the new equipment placing.

The results of our method can be simply interpreted by the specialist because the possible places of access points are defined at the first step.

The practice [14] brings to light the necessity to check the real signal level after theoretical solving of the problem with our method and to implement the method with new measured values. For the system similar to one shown in Fig.1 we have the difference between the real and calculated values of the signal up to 11 dB, the second step gives the maximum error not more than 6 dB which is enough for the engineering calculation.